\documentstyle[11pt,newpasp_frontiers,twoside,epsf]{article}
\def\edcomment#1{\iffalse\marginpar{\raggedright\sl#1\/}\else\relax\fi}
\reversemarginpar

\begin{document}
\title{HD~209458 and the Power of the Dark Side}
  \author{David Charbonneau}
\affil{California Institute of Technology, 105-24, 1200 E. California Blvd., Pasadena, CA 91125}

\begin{abstract}
The rich wealth of observational data, and matching theoretical 
investigations, of the transiting planet of HD~209458 stands
in sharp contrast to systems for which only the radial velocity
orbit is known.  In this paper, I summarize the
current status of these observations, and motivate a variety of projects
that should be accessible with existing instruments.
I describe observational estimates of the planetary radius, and
discuss the relevant sources of uncertainty.  I compare these
estimates to those based on theoretical structural models.
This discussion motivates the observational pursuit of 
three quantities that could be derived from measurements of the 
secondary eclipse:  These are the albedo, the temperature, and the orbital 
eccentricity.  I review the recent detection of the sodium
D lines in the planetary atmosphere, and discuss ongoing work 
to search for molecular features in the near infrared.  I also
outline the use of the Rossiter effect to study the alignment
of the orbit with the stellar equatorial plane, and transit timing
to search for additional objects in the system.
\end{abstract}

\vspace{1cm}

\section{Introduction}
We are fortunate to know of a transiting extrasolar planet
of a nearby, relatively bright star.  Although the method of
photometric transits as a detection technique is still very much 
under development, its application to study the planet of HD~209458 
has proved immensely successful:
HD~209458~b is the only extrasolar planet for which
reliable estimates of the radius and mass are available.
The calculated density proves that the planet is
indeed a gas giant, with a composition primarily of hydrogen
and helium.  In the 2.5 years since its discovery, numerous
observers have seized upon the opportunities afforded by
an extrasolar planet that periodically transits its parent
star.  My goal in this contribution is both to detail the
current state-of-the-art in such observations,
and outline a series of projects that,
with care, should be accessible to current instruments.
This discussion serves to motivate 
wide-field surveys for bright transiting extrasolar planet
systems (see Borucki et al. 2001, Brown \& Charbonneau 2000,
and numerous contributions in this volume).  It is only
for such objects that some of the projects below will
be permitted in the near future.

\section{The Planetary Radius}
\subsection{Observational  Estimates}
Numerous groups (Charbonneau et al. 2000; Deeg, Garrido, \& Claret 2001; 
Henry et al. 2000; Jha et al. 2000) have presented photometric 
observations of the transit with a typical precision of 2~mmag and
a sampling rate of roughly 10~minutes.  At this level of
measurement, attempts to estimate the planetary radius $R_{p}$ 
are frustrated by a significant degeneracy:  Picturing the transit 
as two quantities (a depth and a duration), then it is possible
to fit the data with a family of models, where an 
increase in $R_{p}$ is matched by
a similar increase in the stellar radius $R_{*}$ (thus preserving 
the transit depth), and a decrease in the orbital inclination $i$
(thus preserving the time to transit the chord across
the star).  As a result, these groups used
initial constraints on $R_{*}$ and the stellar mass $M_{*}$, based on
stellar model fits to the observed temperature, brightness,
Hipparcos parallax, and metallicity (Jha et al. 2000; Mazeh et 
al. 2000).  The typical uncertainties in $R_{*}$ 
were $~\sim$10\%, which transfered directly to the uncertainty in $R_{p}$.
As a result, there persisted a significant uncertainty in $R_{p}$,
which was due to limitations in our stellar evolutionary models, 
and specifically our ability to constrain $R_{*}$.

More recently, Brown et al. (2001) observed 4 transits with 
the STIS spectrograph aboard {\it HST}, achieving a photometric 
precision of 0.1~mmag and a cadence of 80~s.  These data broke
the degeneracy described above, since the small changes in the
slope of ingress and egress as a function of $i$ and $R_{*}$
could now be distinguished.  They derived estimates of $R_{p} = 
1.35 \pm 0.06 \, R_{\rm Jup}$ and $R_{*} = 1.15 \pm 0.05 \, R_{\sun}$.
They still needed to assume a value for $M_{*}$ (but not $R_{*}$).  
The degeneracy in the estimate of $R_{p}$ resulting from
the uncertainty in the mass is given by 
\begin{equation}
\frac{\Delta R_{p}}{R_{p}} \simeq 0.3 \, \frac{\Delta M_{*}}{M_{*}}.
\label{eqn1}
\end{equation}
This relation results from the fact that, for model fits 
to an observed transit curve, $R_{p} \propto R_{*} \simeq v_{orb} \ \Theta_{\rm I}$, where
$v_{orb}$ is the planetary orbital velocity and $\Theta_{\rm I}$ is the
transit duration.
The quantity $\Theta_{\rm I}$ is observed (and thus fixed), and 
$v_{orb} \propto M_{*}^{1/3}$.
Equation (1) results in the limit of small $\Delta M_{*}/M_{*}$.

Cody \& Sasselov (2002) have undertaken a detailed study of 
stellar evolutionary
models of HD~209458 and the resulting uncertainties in $R_{p}$.  
In particular, they point out that the mass-radius
relation at constant luminosity yields the dependence
\begin{equation}
\frac{\Delta R_{p}}{R_{p}} \simeq - \, \frac{\Delta M_{*}}{M_{*}},
\end{equation}
which is nearly orthogonal to that resulting from light-curve fitting
(equation 1).  They 
recommend explicitly assuming the relationship between $R_{*}$ and $M_{*}$ 
resulting from model fitting, rather than independently
considering each with uncorrelated uncertainties.  The result
should be to reduce the net effective uncertainty in the estimate of
$R_{p}$.

\subsection{Theoretical Understanding}
The inflated size of $R_{p}$ relative to
our own Jupiter was predicted by Guillot et al. (1996)
prior to the discovery of transits in the HD~209458 system.
Burrows et al. (2000) point out that the large radius 
results from the inability of the planet to cool
in the hot environment resulting from the intense
stellar insolation:  The stellar flux incident upon
HD~209458~b is roughly 20,000 times that received at
Jupiter.  They find that the planet
must have migrated inwards to its current location
in less than $\sim$10~Myr after its formation.  Guillot \& Showman (2002) 
re-examine the sources of opacity in the planetary atmosphere,
and find that the models of Burrows et al. (2000) may
have deposited the stellar radiation too
deep in the atmosphere.  They calculate
cooler model atmospheres, and predict a value for 
$R_{p}$ significantly below the observed value.  The inclusion of a
core of high-density material (Bodenheimer, Lin, \& Mardling 2001) increases
the disparity.

The realization that the observed $R_{p}$ may be significantly
larger than that predicted from theoretical calculations 
has lead several groups to consider
additional sources of energy in the bulk of the planet.  
Bodenheimer et al. (2001) 
show that tidal dissipation of orbital eccentricity could
provide the required input.  However, the orbit would be 
circularized on a time scale of $\sim$10$^{8}$~yr, and thus a mechanism
to excite the orbital eccentricity would also be required.
Showman \& Guillot (2002) show that terrific winds may be
expected on HD~209458~b, and describe a model
that allows for a downward transport of kinetic energy
of $\sim$1\% of the incident stellar flux, sufficient to
maintain the observed $R_{p}$.

The contribution of observers in this debate is to better
quantify the relevant observables.  Three quantities of interest
are (1) the actual temperature of the planet, (2) the fraction 
of stellar flux absorbed by the planet, and 
(3) additional sources of energy (in the bulk of the planet) 
beyond the effect of insolation.  
In the next section, I describe how observations of
the secondary eclipse (the passage of the planet behind the
star) will allow estimates of these quantities, should the
requisite precision be achieved.

\section{Pursuing the Secondary Eclipse}
\subsection{Reflected Light}
Planets shine in reflected light with a flux (relative
to their stars) of
\begin{equation}
\left( \frac{f_{p}}{f_{*}} \right)_{\lambda} (\alpha) = \left( \frac{R_p}{a} \right )^2 \, p_{\lambda} \, \Phi_{\lambda}(\alpha),
\end{equation}
where $a$ is the semi-major axis, $p_{\lambda}$ is the
geometric albedo, and $\Phi_{\lambda}(\alpha)$ is the phase
function (the flux from the planet when viewed at a phase
angle $\alpha$ relative to the flux received when the
planet is at opposition).  Charbonneau et al. (1999) and 
Collier Cameron et al. (2001) have presented upper limits on 
$p_{\lambda}$ for the hot Jupiter orbiting
$\tau$~Boo, and Collier Cameron et al. (2002) have presented a
similar study of the innermost planet of Ups~And.
Since these planets do not transit, the authors assumed 
values for the $R_p$, $\alpha$ (a function of $i$), and 
$\Phi_{\lambda}(\alpha)$,
which complicated the interpretation of their results.

The situation is greatly simplified in the case of HD~209458,
since $R_{p}$ and $i$ are known with high accuracy 
(Charbonneau \& Noyes 2000). Furthermore,
near the time of secondary eclipse, the planet seen just
before ingress or just after egress is only $\sim$6${\deg}$
from opposition.  Under the approximation $\Phi (6{\deg}) 
\simeq \Phi(0{\deg}) \equiv 1$, equation (3) yields 
the prediction for the depth of secondary eclipse,
\begin{equation}
\left( \frac{\Delta f}{f} \right)_{\lambda} \simeq 2.0 \times 10^{-4} \, p_{\lambda}.
\end{equation}
Thus $p_{\lambda}$ can be measured directly. 
Brown et al. (2001) demonstrated that STIS should be able to achieve 
the required precision.

The quantity $p_{\lambda}$ is one element in 
an estimate of the total energy deposited into
the planetary atmosphere.  This is expressed as $1-A$, where
$A$ is the Bond albedo.  Converting from
$p_{\lambda}$ to $A$ requires both $p_{\lambda}$ over the 
wavelength range where the star outputs most of its energy, 
and an estimate of 
$\Phi_{\lambda} (\alpha)$.  A theoretical investigation by 
Seager, Whitney, \& Sasselov (2000) 
shows that a wide variety of phase functions are
possible.  The Canadian MOST satellite (Matthews et al. 2000) 
will attempt to measure the phase variability of several
hot Jupiters.  In addition to providing the last step in an 
estimate of $A$, the measurement of $\Phi_{\lambda}(\alpha)$ is highly 
diagnostic of sources of scattering in the atmosphere.

\subsection{Thermal Emission}
Infrared photometry of the secondary eclipse offers
the opportunity to estimate the planetary temperature.
In the Rayleigh-Jeans limit, the depth of the secondary
eclipse is given by the product of the observed depth of the
primary eclipse (in the absence of limb-darkening) and
the ratio of the object temperatures,
\begin{equation}
\frac{\Delta f}{f} \simeq \frac{T_{p}}{T_{*}} \, \left ( \frac{R_{p}}{R_{*}} \right )^2 \simeq \frac{1500 \, {\rm K}}{6000 \, {\rm K}} \, 0.0146 \simeq 4 \, {\rm mmag}. 
\end{equation}
This signal is roughly 20 times that of equation (4).
In practice, such observations are frustrated by bright
thermal background levels:  Infrared arrays
tend to cover small areas of the sky (to avoid saturation
from background thermal emission), which means that the 
bright calibration stars required for mmag-precision differential 
photometry are not available.  The technique of slewing 
to nearby bright stars
suffers from significant changes in extinction which occur
even on short time scales at these long wavelengths.

An alternative approach is spectroscopy.  Richardson et al. (2003)
have analyzed VLT spectra with $R=3300$ at 3.5--3.7~$\mu$m to search
for the disappearance of planetary methane features at the
time of secondary eclipse.  They are currently able to exclude
a hot ($A=0$) model at roughly the 2$\sigma$ level.

\subsection{Effects of Orbital Eccentricity}
The current upper limit on the orbital eccentricity $e$ from
radial velocity measurements is consistent with zero 
($e = 0.00967 \pm 0.014$; G.~Marcy, personal communication).
Theoretical considerations (Goldreich \& Soter 1966) predict that 
the orbit should be circularized on a time scale of 
$\tau_{e} \simeq 3 \times 10^{7} \, (Q/10^{5})$~yr 
(where $Q$ is the tidal quality factor).  Indeed, all
known hot Jupiters with orbital periods $P$ less than 5~days are
observed to be circular.  

A non-zero value of $e$ could
produce a measurable shift in the separation of the times
of the center of primary ($t_{\rm I}$) and secondary ($t_{\rm II}$) eclipse
away from a half-period.  The approximate formula for the timing offset 
(see Kallrath \& Milone 1998 and references therein) is:
\begin{equation}
\frac{\pi}{2 \, P} \left ( t_{\rm II} - t_{\rm I} - \frac{P}{2} \right ) \simeq e \, \cos \omega \le e,
\end{equation}
where $\omega$ is the longitude of periastron.
At the $2\sigma$ limit on the current uncertainty in $e$, 
the time of secondary eclipse could be offset by as much as 2.0~hrs,
a shift that should be readily detectable if the precision to observe
the secondary eclipse is obtained.  In the case of a detection, 
equation (6) provides a lower limit on the true value of the eccentricity
(and the radial velocity yields an upper limit on its value).
Should no offset be seen, a significant upper limit would 
still be of interest, as it would rule out heating of the planet 
interior by tidal damping of the orbital eccentricity.

An eccentric orbit could also change the relative durations of the
primary ($\Theta_{\rm I}$) and secondary ($\Theta_{\rm II}$) eclipses
(again, see Kallrath \& Milone 1998).  The relevant formula is:
\begin{equation}
\frac{\Theta_{\rm I} - \Theta_{\rm II}}{\Theta_{\rm I} + \Theta_{\rm II}} \simeq e \, \sin \omega.
\end{equation}
At the limit of the $2\sigma$ error bars on $e$, 
this yields a change of 14~minutes,
a more ambitious measurement that the offset of equation (6).  
Nonetheless, equations (6) \& (7) together allow for 
the direct evaluation of $e$ and $\omega$.

\section{Atmospheric Transmission Spectroscopy}
Shortly after the detection of the photometric transits of HD~209458~b,
several groups (Seager \& Sasselov 2000; Brown 2001; Hubbard et al. 2001) 
presented theoretical transmission spectra of the planetary atmosphere.
The essential idea is that opacity sources in the atmosphere result
in a wavelength dependence of the apparent radius as derived from 
transit observations.  Thus, examining the ratio of stellar
spectra taken in and out of transit may yield a probe
of the dominant sources of opacity in the planetary atmosphere.

The large equilibrium temperature of the
planet ($\sim$1500~K) implies a scale height of 550~km.  Thus
the atmosphere presents an annulus with a 
one-scale-height cross-sectional area (relative to the star)
of $\Delta A / A = 2.6 \times 10^{-4}$.  Some features may have effective
depths of several scale heights, and thus signals as large
as $1 \times 10^{-3}$ could be produced.

\subsection{Detection of the Sodium D Lines}
Using STIS, Charbonneau et al. (2002) have detected an increase
in the transit depth of $(2.32 \pm 0.57) \times 10^{-4}$ in
a 1.2~nm bandpass centered on the sodium D lines (located
at 589.3~nm) relative to the local continuum.  
They rule out alternate sources for this
decrement (most notably stellar limb darkening), and consider
implications for the planetary atmosphere.  They find that
the signal is roughly a factor of 3 smaller than 
their predictions for a cloudless model of the atmosphere with
a solar abundance of sodium in atomic form.  Although the
interpretation of this result is far from unique, they
quantify possible models to explain the observed signal:
If the disparity between the signal and their fiducial model is due 
entirely to clouds, then a very high cloud deck (with cloud tops
above 0.4~mbar) is required.  Alternately, it may be that 
the atmosphere is depleted in atomic sodium:  Models that
leave less that 1\% of a solar abundance of sodium
in atomic form are capable of reproducing the results.

Since the detection, several groups have revisited models
of the planetary atmosphere.  Barman et al. (2002) consider non-LTE
effects.  Fortney et al. (2002)
conduct a detailed examination of ionization and
cloud formation, and are able to produce models that
lie within the error bars of the detection.

\subsection{Infrared Observations}
Although sodium is spectroscopically very active, it is only a trace
constituent of the planetary atmosphere.  Molecules such as
H$_2$O, CH$_4$, and CO are of much greater diagnostic potential.
The molecule CO is of particular interest:  The equilibrium
temperature of HD~209458~b ($\sim$1500~K) happens to fall
in the narrow regime where the dominant carbon bearing molecule
switches from CO to CH$_4$, with CO preferred in the hotter state.
A search for CO is aided by the lack of this feature in
the telluric spectrum (unlike CH$_4$).  

Brown, Libbrecht, \& Charbonneau (2002) presented data from
an exploratory night on HD~209458 with the NIRSPEC instrument
on Keck~II.  The observing conditions were not ideal:  
They were required to use the adaptive optics system,
the weather was poor, and the transit was not entirely visible
from their longitude.  All of these factors served to reduce
the number of photons that they gathered.  They presented an upper 
limit that was roughly a factor of 3 too great to test realistic models
of the planetary atmosphere.  Nonetheless, it is reasonable
to suppose that great gains in the photon-noise limited signal-to-noise 
ratio (SNR) are possible by observing a couple transits centered
near stellar meridian passage and under excellent weather conditions.

At high resolution, the orbital velocity of the planet must
be taken into account.  The planet's radial velocity shifts
by roughly 30~\mbox{km s$^{-1}$} during the course
of the transit.  The detection of this change in
the Doppler shift would provide convincing evidence that
a candidate signal is planetary in origin.

\section{Additional Effects During Transit}
\subsection{Rossiter Effect}
Queloz et al. (2000) and Bundy \& Marcy (2000) have presented
radial velocities during transit.  After subracting
the known orbit, they find an initial redward, then blueward 
swing, caused by the planet's
occultation of the approaching and receding limbs of
the rotating star.  This effect is called the Rossiter
effect, after Rossiter (1924).  These data confirm
that the planet orbits in the same sense as the star
rotates.  Queloz et al. (2000) model their data in
detail to show that the planetary orbit appears to
be co-aligned with the apparent stellar equatorial
plane.  They calculate that the 
time scale for this alignment to result from the tidal
influence of the planet upon the star is 
significantly longer than the age of the system.  This indicates 
that the alignment is primordial, as would be expected for formation 
and migration of the planet in a protoplanetary disk.  
However, their uncertainties on the angle between the
orbital plane and the stellar equatorial plane are very large: 
It would be of considerable interest to repeat their
experiment and see if the relative alignment of the 
axes is less than several degrees (as is the case for the solar system).
Furthermore, it should be possible to measure the actual distortion
to the stellar line profiles (caused by the planetary occultation) 
in high-resolution, high-SNR spectra.  Modeling these distortions
should provide a radial velocity map of the star at the 
latitude of the transit.  To date, only the net Doppler
shift of the lines have been reported. 

\subsection{Transit Timing}
Additional objects in the HD~209458 system may be revealed
by variations in the observed times of center of transit ($T_{c}$)
relative to the predictions of a simple orbital period.
Brown et al. (2001) used STIS observations to rule out a planetary
satellite with a mass in excess of 3~$M_{\earth}$ (they also 
placed limits on the physical radius of a satellite).
More recently, Schultz et al. (2003) have presented observations
of 4 transits with the Fine Guidance Sensors (FGS) aboard {\it HST}.  
The FGS provided photometry with a SNR ratio of $\sim$80 and
a very rapid sampling rate (0.025~s).  Their observations targeted the
times of ingress and egress.  In addition to searching for 
planetary satellites or additional planets, these data can be combined 
with those of Brown et al. (2001) to derive an extremely precise 
estimate of the planetary orbital period.  This period will
be of great practical value since it will effectively remove 
timing uncertainties in planning and interpreting
observations of future transits.

\acknowledgements I wish to thank Tim~Brown for his
insight into this crazy, mixed-up world.

\end{document}